\documentclass{cmspaper}

\usepackage{graphicx}

\usepackage{amsmath}

\usepackage[latin1]{inputenc}
\usepackage[amssymb,thinspace]{SIunits}
\usepackage{wasysym}

\newcommand{\MeV}{\mega\electronvolt}
\newcommand{\cm}{\centi\meter}
\newcommand{\mm}{\milli\meter}
\newcommand{\mum}{\micro\meter}
\newcommand{\nm}{\nano\meter}

\begin{document}

\graphicspath{{fig/}}


\begin{titlepage}

   \date{\today}

   \title{Validation of Kalman Filter alignment algorithm with cosmic-ray data using a CMS silicon strip tracker endcap}

  \begin{Authlist}
    D. Sprenger, M. Weber\footnote{Now at University of Hamburg, Hamburg, Germany},
    R. Adolphi, R. Brauer, L. Feld, K. Klein, A. Ostapchuk, S. Schael, B. Wittmer 
    \Instfoot{rwth}{RWTH Aachen University, I. Physikalisches Institut, Aachen, Germany}

  \end{Authlist}

  \begin{abstract}
    A Kalman Filter alignment algorithm has been applied to cosmic-ray data.
    We discuss the alignment algorithm and an experiment-independent
    implementation including outlier rejection and
    treatment of weakly determined parameters. Using this implementation, the
    algorithm has been applied to data recorded with one CMS silicon tracker
    endcap. Results are compared to both photogrammetry measurements and data
    obtained from a dedicated hardware alignment system, and good agreement is
    observed.
  \end{abstract}

\end{titlepage}

\setcounter{page}{2}


\section{Introduction}
\label{sec:introduction}

Todays tracking detectors in particle physics experiments consist of several
hundreds up to ten thousands of independent detector elements, which allow to
measure charged particle trajectories with a single-point resolution of
typically $10$--$50\,\mum$~\cite{babar,atlas,cms}. This resolution is
significantly better than the placement accuracy as achieved during
construction, which typically is an order of magnitude larger. The established
method to determine the true detector element position and orientation is to
use measurements from particles traversing the detector in-situ, minimizing
the residuals of an appropriate track model. Several algorithms have been
proposed and exposed to data~\cite{millepede,babar-alignment}. In this note
we discuss the implementation and application of an alignment algorithm
based on the Kalman Filter~\cite{kalman-jphysg} to tracks from cosmic muons
recorded by an integration setup of one CMS Tracker
Endcap~\cite{endcap-integration}, detailing the algorithm and summarizing
results from reference~\cite{sprenger}. The algorithmic implementation is done in a
portable, experiment-independent manner and easily allows application in other
experiments as well.


\section{Kalman Filter alignment algorithm}
\label{sec:kalmanalgorithm}

The Kalman Filter algorithm has been implemented mainly according to the
formul\ae\ described in reference~\cite{kalman-jphysg} with some modifications.

The vector $\vec m$ of detector measurements on a particle track is described
by a function $\vec f$ that depends both on the track parameters $\vec p$ and
the alignment parameters $\vec a$,
\begin{equation}
  \vec{m} = \vec f (\vec p, \vec a) + \vec \epsilon,
\end{equation}
where the measurement errors are described by $\vec \epsilon$, which has a
known covariance matrix $C$. If the function $\vec f$ is not linear in the
parameters $(\vec p, \vec a)$, it is linearized at a starting point $(\vec
p_0, \vec a_0)$:
\begin{eqnarray}
  \vec m &=& \vec f (\vec p_0, \vec a_0) + H (\vec p - \vec p_0) + D (\vec a -
  \vec a_0) + \vec \epsilon + \mathcal{O}\left((\vec p - \vec p_0)^2, (\vec a -
  \vec a_0)^2\right)\\
         &\approx& \underbrace{\vec f (\vec p_0, \vec a_0) - H \vec p_0 -
                   D \vec a_0}_{\vec c} + H \vec p + D \vec a 
  \label{eqn:kalmanalgorithm:linearization} 
\end{eqnarray}
where the Jacobians $H$ and $D$ are
\begin{equation}
  H = \frac{\partial \vec f}{\partial \vec p}(\vec p_0, \vec a_0),\qquad  D
  = \frac{\partial \vec f}{\partial \vec a}(\vec p_0, \vec a_0).
  \label{eqn:kalmanalgorithm:jacobians}
\end{equation}
Typically, for the expansion point $\vec a_0$ the design geometry, knowledge
from the assembly, or a previous alignment is chosen, and $\vec p_0$ are the track
parameters obtained with this geometry.

The goal of the Kalman Filter algorithm is to minimize the track residuals,
i.\,e.\ to minimize the objective function 
\begin{equation}
  f_{obj} = (\vec m - \vec f(\vec p, \vec a))^T C (\vec m - \vec f(\vec
  p, \vec a)) \label{eqn:kalmanalgorithm:chi2}
\end{equation}
for the given track sample. This is achieved processing tracks in sequence and
updating parameters and covariance matrix after each track. The resulting
update equations for the alignment parameters $\vec a$ and their covariance
matrix $E$ are~\cite{kalman-jphysg,cms-note-2006-022}
\begin{eqnarray}
  \hat{ \vec a} &=& \vec a + E D^T W  (\vec m - \vec c - D \vec a)
  \label{eqn:kalmanalgorithm:updateparameters}\\
  \hat E &=& E - (E D^T)^T \cdot W \cdot E D^T
  \label{eqn:kalmanalgorithm:updatecovariance}
\end{eqnarray}

Here, $W$ and $V$ are auxiliary matrices, given by~\cite{cms-note-2006-022}
\begin{equation}
  W = V^{-1} - V^{-1} H (H^T V^{-1} H)^{-1} H^T V^{-1}, \qquad V = C + D E
  D^T.
 \label{eqn:kalmanalgorithm:auxiliary}
\end{equation}


Alignment parameters $\vec a$ and covariance matrix $E$ are updated with each
track. The algorithm needs a starting point for the parameters $\vec a$ and
their covariance matrix $E$, which can be set to expectations e.\,g.\ from
assembly tolerances. If in doubt, larger initial uncertainties are preferred
since too small values can bias the alignment because the covariance matrix
has decreasing eigenvalues. 
Parameters corresponding to global degrees of freedom can be fixed by
assigning tiny prior uncertainties to specific alignment parameters.

In order to avoid bias from mis-measured tracks in the first steps,
deterministic annealing with a configurable geometric schedule can be
applied. This is done by increasing measurement uncertainties in
Equation~(\ref{eqn:kalmanalgorithm:auxiliary}): 
\begin{equation}
  V = \alpha(k) C + D E D^T, \qquad \alpha(k) = \left\{ 
  \begin{array}{ll}
  b^{\frac{n-k}{n-1}} & \mbox{for } 1 \le k \le n, \\
  1                   & \mbox{for } k > n
  \end{array}
  \right.
\end{equation}
Here, $k$ denotes the number of the current track, $b$ is the annealing factor
which is applied to the covariance matrix in the first step, and $n$ is the
number of the track from which on the full information from the track is
kept.


\section{Implementation}
\label{sec:implementation}

The software implementation of the algorithm has been done in an
experiment-independent way. However, the input data are obtained from a specialized
software framework tailored to the experiment, usually providing pattern
recognition, track reconstruction as well as the description of the
geometrical layout of the detector elements. Therefore, an interface between
the experiment specific and the experiment independent software was designed.
The basic choice was made that the experiment specific implementation has to
provide all the information that is necessary to compute
Equations~(\ref{eqn:kalmanalgorithm:updateparameters})--(\ref{eqn:kalmanalgorithm:auxiliary})
in a persistent matrix format, plus some additional information.

\subsection{Experiment specific implementation}
\label{sec:implementation:experiment-specific}

In the experiment specific implementation, the first additional information to
be provided and stored is the number of parameters to be aligned. Parameters
need to be identified uniquely by an index. This is necessary to save space
and speed up the computation. The reason is that the Jacobian $D$ in
Eq.~(\ref{eqn:kalmanalgorithm:jacobians}) has only few entries different from
zero, since a single track only crosses few detector elements.

A rough selection of tracks suitable for alignment has to be done, judging on
the current alignment parameters. Then, for each track, the experiment
specific software has to supply the measurement $\vec m$ with its covariance
$C$, the constant $\vec c$, and the Jacobians $H$ and $D$. The Jacobians are
evaluated according to the current knowledge at non-optimal parameters $\vec
p_0$ and $\vec a_0$
\footnote{A reference implementation in C++~\cite{cpp} for the CMS
experiment is available and can be used as a template for other experiments.}.

Additional experiment dependent information can be supplied, like run and
event numbers and the number of the track in the current event. This can be
especially useful if some tracks are rejected as outliers in the alignment
procedure, such that one can have a look at the corresponding
events within the experiment dependent software (e.\,g.\ event display). Also
the value $\tilde\chi^2$ of the track fit, defined as
\begin{equation}
  \tilde\chi^2 = \vec{r}{\,}^T\,C \,\vec r \qquad \mbox{with}\qquad \vec r =
  \left(\vec m - \vec c\right) \label{eqn:implementation:chi2}
\end{equation}
and the number of degrees of freedom $n_{dof}$ in the fit are stored. One has
to note that in the case of an unaligned detector, $\tilde\chi^2$ does not
follow a $\chi^2$-distribution, but still provides some power to discriminate
bad tracks. 

\subsection{Implementation of the Kalman Filter alignment algorithm}
\label{sec:implementation:kalmanfilter}

The Kalman Filter alignment algorithm is implemented as a C++ program, which
uses the ROOT~\cite{root} data analysis framework. The program reads the
needed track information from the files which were created with the
experiment specific program.

The program initializes a vector of alignment parameters and the corresponding
covariance matrix with configurable starting values. One option is to pass
over this information from the experiment specific part. However,
pattern recognition and track reconstruction are very costly time-wise, and
input values to alignment only change little (e.\,g.\ due to non-linearities
of $\vec f (\vec p, \vec a)$, which are neglected in
Equation~(\ref{eqn:kalmanalgorithm:linearization})) when a different starting
point is taken.
By choosing the initialization to happen in the experiment independent part,
avoiding a new reconstruction, computing time is saved.

Then, processing track after track, the parameters $\vec a$ and the covariance
matrix $E$ are updated as specified in
Equations~(\ref{eqn:kalmanalgorithm:updateparameters}) and
(\ref{eqn:kalmanalgorithm:updatecovariance}). 
Memory and computing time are saved in this step by reducing
the alignment derivatives matrix $D$ to non-zero columns, which correspond to
the detector elements hit by the current track. In the same spirit, computing
time is saved by using only the (known) non-zero elements of $D^T$ when
computing the product $ED^T$. The matrix $ED^T$ consists of one line for each
alignment parameter (not only those hit by the current track), and one column
for each single measurement of the current track. The most computing time
intensive matrix operation is the update of the matrix $E$, which is quadratic
in the number of parameters and dominates the time consumption of the
alignment procedure in case of many alignment parameters.

During the update, the current alignment parameters and their uncertainties
(extracted from the diagonal elements of the covariance matrix) are filled into
histograms. After all tracks have been processed, these histograms together with
the final parameters, their uncertainties, and additional information on the alignment
procedure like the number of hits for each detector element are stored.

\subsubsection{Outlier rejection}
\label{sec:implementation:kalmanfilter:outlierrejection}

To prevent bad measurements (e.\,g.\ from noise clusters) from having a large
impact on the alignment results, outlier rejection is implemented. As
mentioned above, the value $\tilde\chi^2$ (cf.\
Eq.~(\ref{eqn:implementation:chi2})) is computed for each track. The track is
used for updating $\vec a$ and $E$ when the probability $P(\tilde\chi^2,
n_{dof})$ exceeds a configurable threshold value.

An additional outlier rejection is implemented by rejecting tracks that change
the alignment parameters by more than a configurable factor beyond their
uncertainties. Tracks showing this effect are assumed to be wrongly measured.
Considering the fact that the current alignment parameters
combine the information of all previously processed tracks, one additional
track should not have such a large impact on the results.

\subsubsection{Weakly determined parameters}
\label{sec:kalmanalgorithm:coordinatesystem}

The objective function (\ref{eqn:kalmanalgorithm:chi2}) is invariant under
a global translation and rotation of all detector elements, i.\,e.\ a special
linear transformation of the alignment parameters. Even when not present in
the starting values $\vec a$ and $E$, such transformations can build up due to
mis-measurements, round-off problems in the alignment procedure, or an
incomplete track model. Large alignment parameters $\vec a$ have a negative
impact on the alignment procedure due to larger distance to the linearization
point. Therefore, these global parameters should be fixed. This can be achieved by
assigning small uncertainties to some (linear combinations of) parameters for
the initial matrix $E$, depending on the experimental setup, which corresponds
to the definition of a geometry reference system.

Furthermore, large eigenvalues can be present in $E$ even after processing the
last track. These large eigenvalues correspond to certain linear combinations
of alignment parameters which geometrically represent a systematic distortion
of the detector units position and orientation that is only weakly
determined. This happens especially if the track sample consists of tracks
with similar topology. Artificial distortions can bias physics observables
like invariant masses, momentum scale etc.

We consider two methods to suppress weak modes: The first method is to assign
a small initial error to the weakly determined linear combination of
parameters in $E$. The alternative is to align without changing the initial
values of $E$, and later fit the amplitude of the weakly determined mode and
subtract it from the parameters.


\section{Application to data}
\label{sec:application}

The Kalman Filter alignment algorithm was applied to data taken from the
integration setup of one endcap of the CMS
experiment~\cite{cms,endcap-integration} tracker (TEC+). The integration took
place in 2006 at RWTH Aachen University. Apart from commissioning the system
hardware, tracks from cosmic muons originating from air-showers were recorded
and used for alignment.

\subsection{CMS Tracker Endcap}
\label{sec:application:cmstracker:tec}

\begin{figure}[ht]
  \begin{center}
    \includegraphics[width=0.9\textwidth]{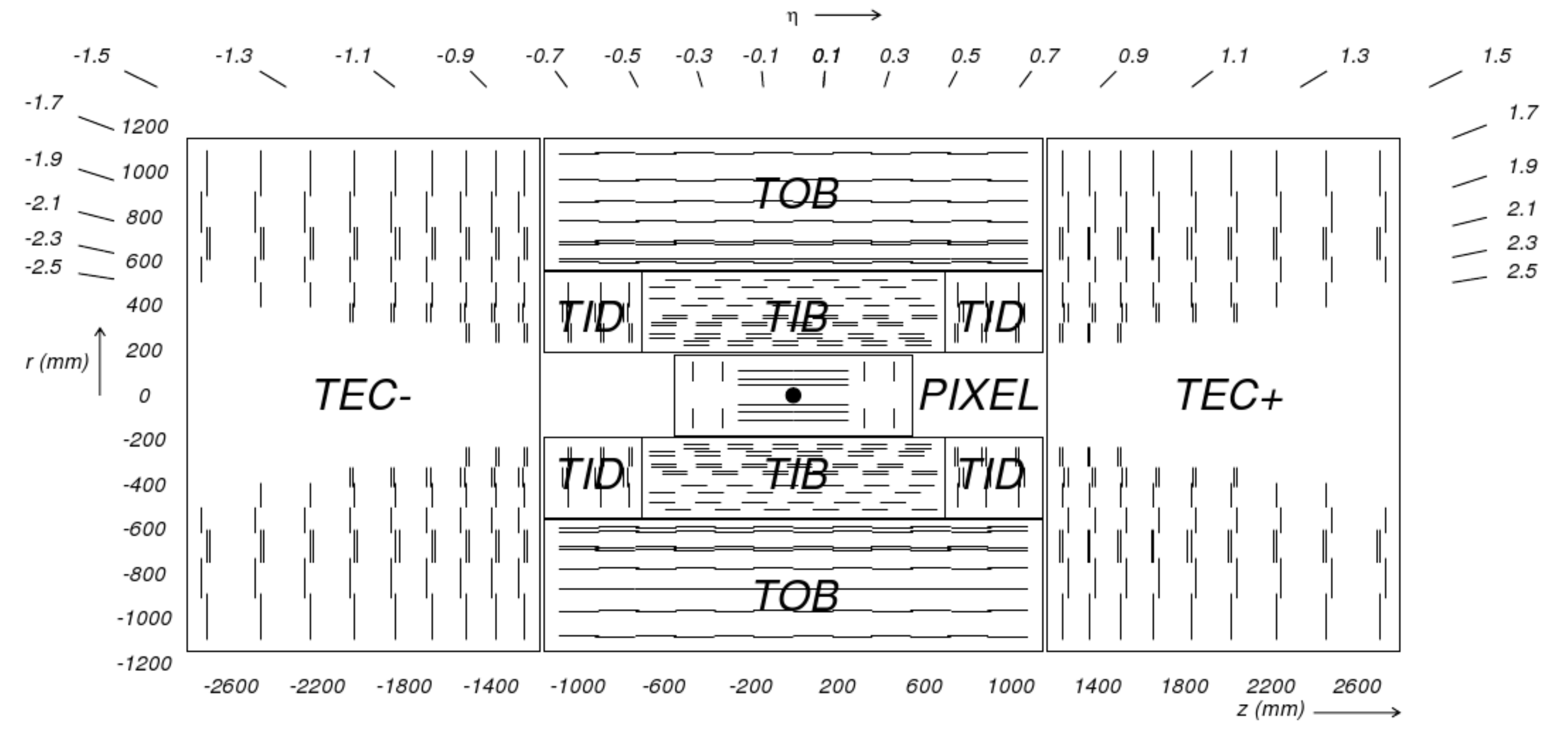}
    \caption{View of the CMS tracker in the
    $rz$-plane~\cite{cms}. Each line in the strip tracker represents a silicon
    strip detector, whereas lines in the pixel detector represent ladders and
    petals on which the detectors are mounted in the barrel and endcaps,
    respectively.}
    \label{fig:application:cmstracker:cmstracker}
  \end{center}
\end{figure}

The CMS tracker~\cite{cms} is entirely based on silicon detector
technology. It can be divided into five subsystems
(Fig.~\ref{fig:application:cmstracker:cmstracker}): Pixel detector (PIXEL),
Tracker Inner Barrel (TIB), Tracker Outer Barrel (TOB), Tracker Inner Disks
(TID+, TID-), and Tracker Endcaps (TEC+, TEC-).

CMS uses the following coordinate definitions: The $y$-axis points upwards,
the $x$-axis points radially to the centre of the LHC~\cite{lhc} ring, and the
$z$-axis points in direction of the beam line, completing a right-handed
coordinate system. The azimuthal angle $\phi$ is measured to the $x$-axis in
the $x$-$y$-plane, and hence describes rotations around the $z$-axis, and the
polar angle $\theta$ is measured to the $z$-axis.
 
\begin{figure}[ht]
  \begin{center}
    \includegraphics[width=0.3\textwidth]{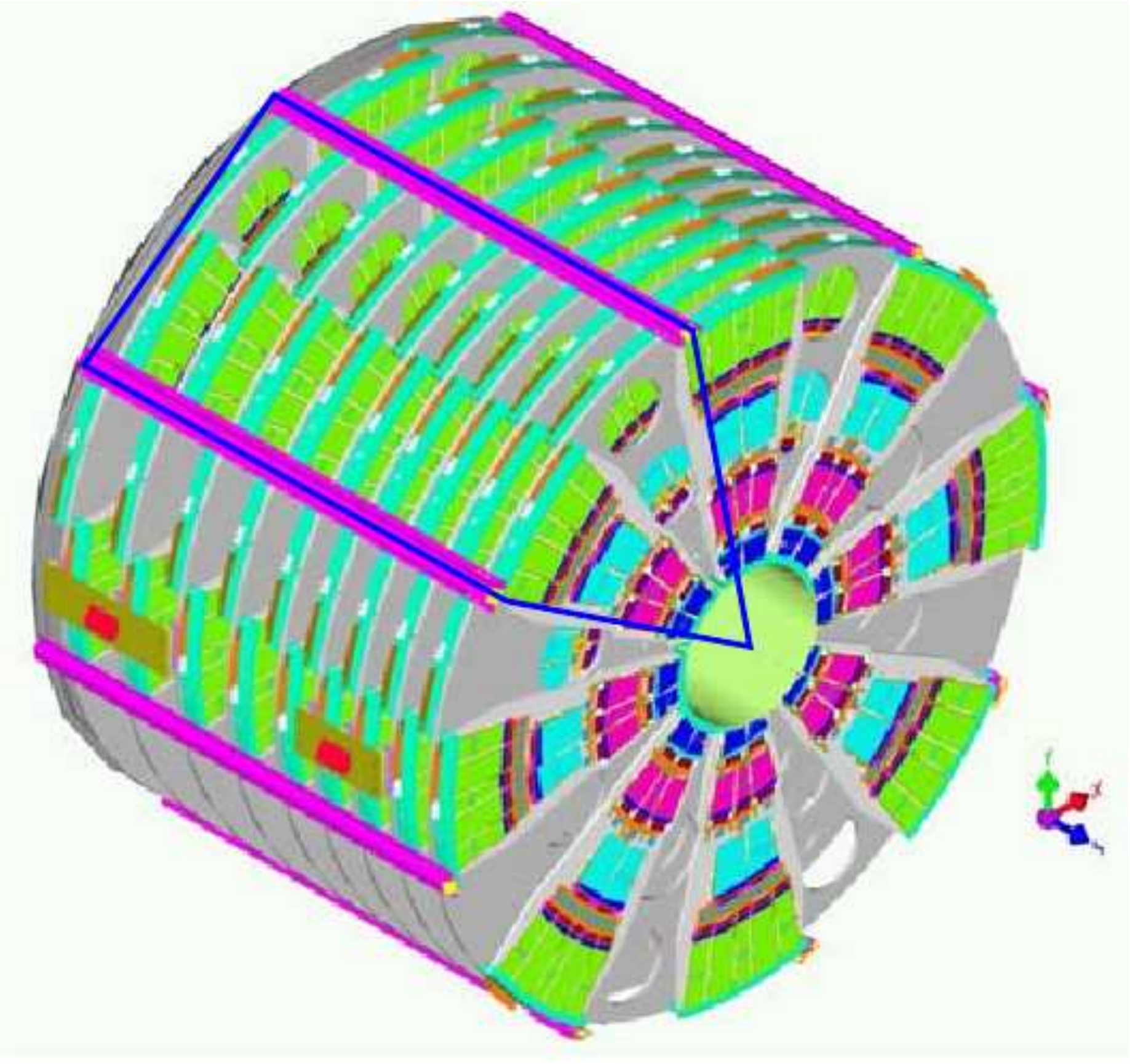}\hfill
    \raisebox{3mm}{\includegraphics[width=0.25\textwidth]{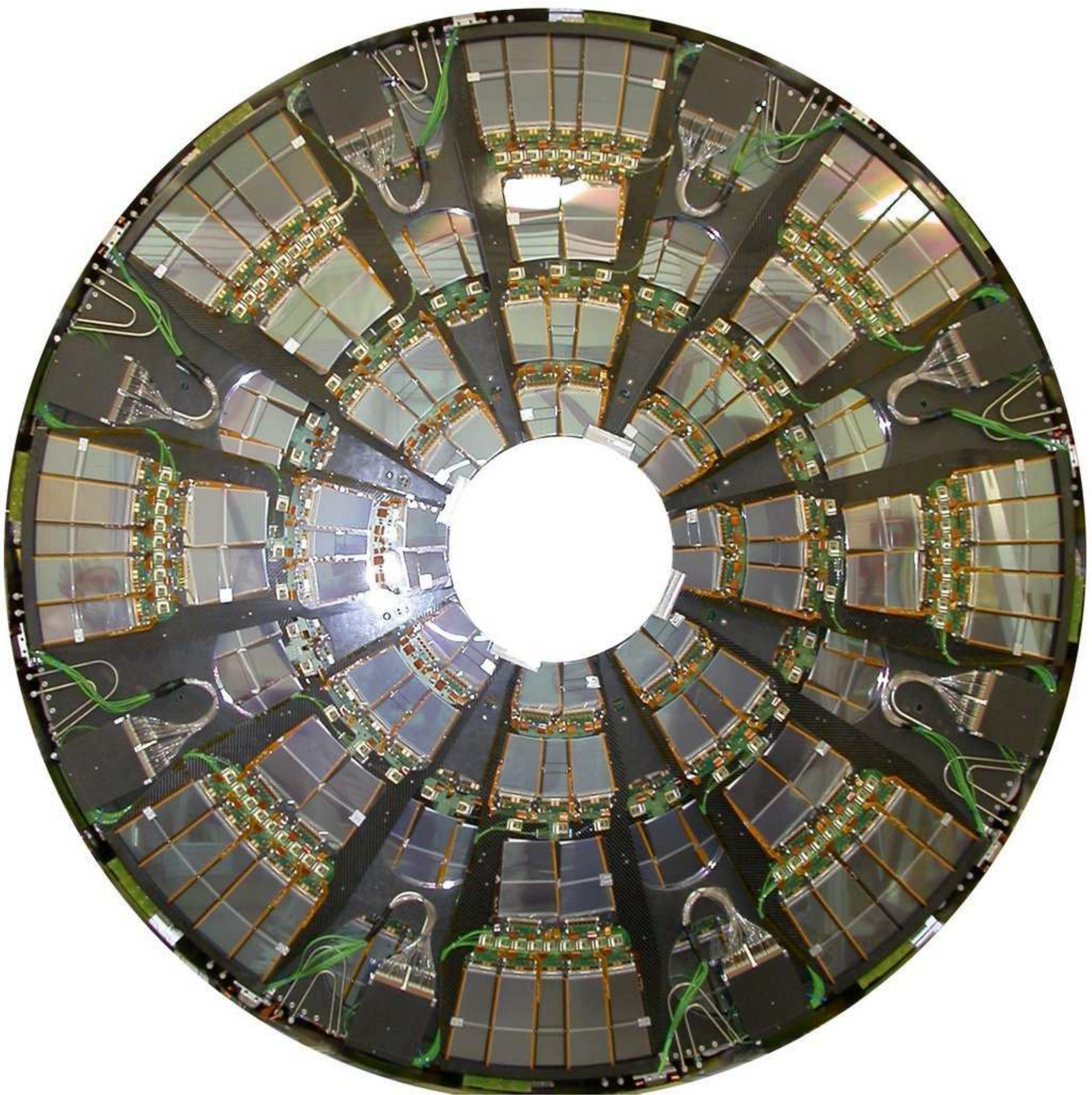}}\hfill
    \includegraphics[width=0.3\textwidth]{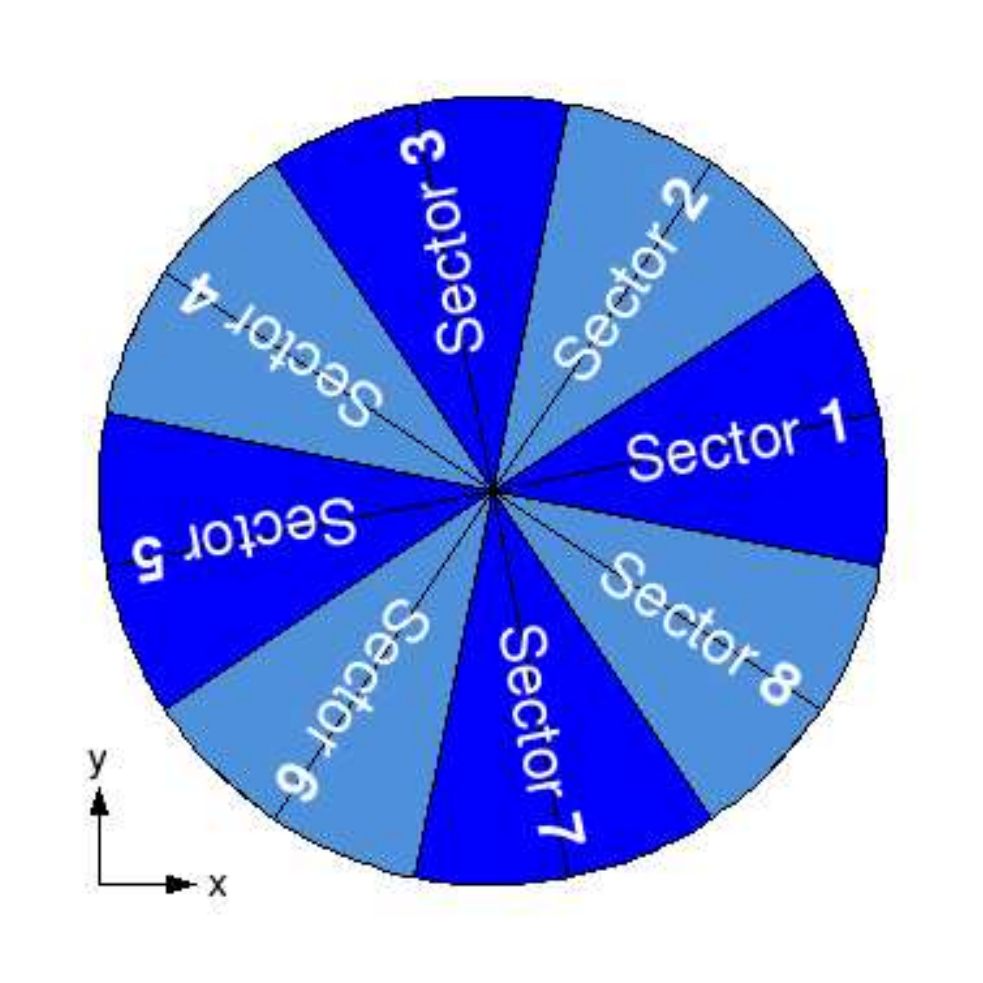}
    \caption{Left: Sketch of a tracker endcap~\cite{cms}. Middle:
    Photo of a TEC disk. Right: Sector numbering scheme.}
    \label{fig:application:cmstracker:tec:endcapdisk}
  \end{center}
\end{figure}

\begin{figure}[ht]
  \begin{center}
    \begin{minipage}{6.0cm}
      \centerline{\includegraphics[width=6.0cm]{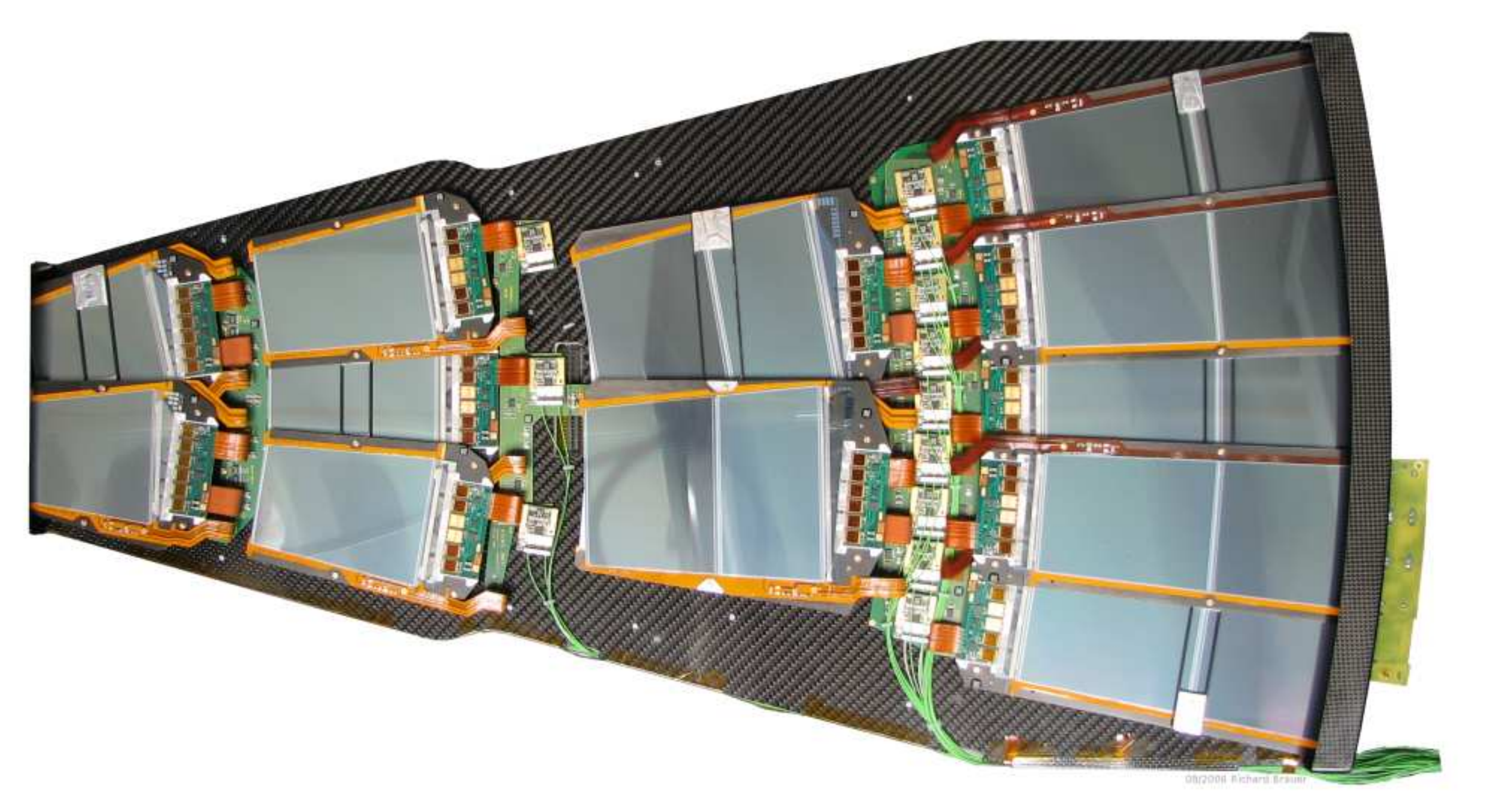}}
    \end{minipage}
    \hspace{0.6cm}
    \begin{minipage}{6.0cm}
      \centerline{\includegraphics[width=6.0cm]{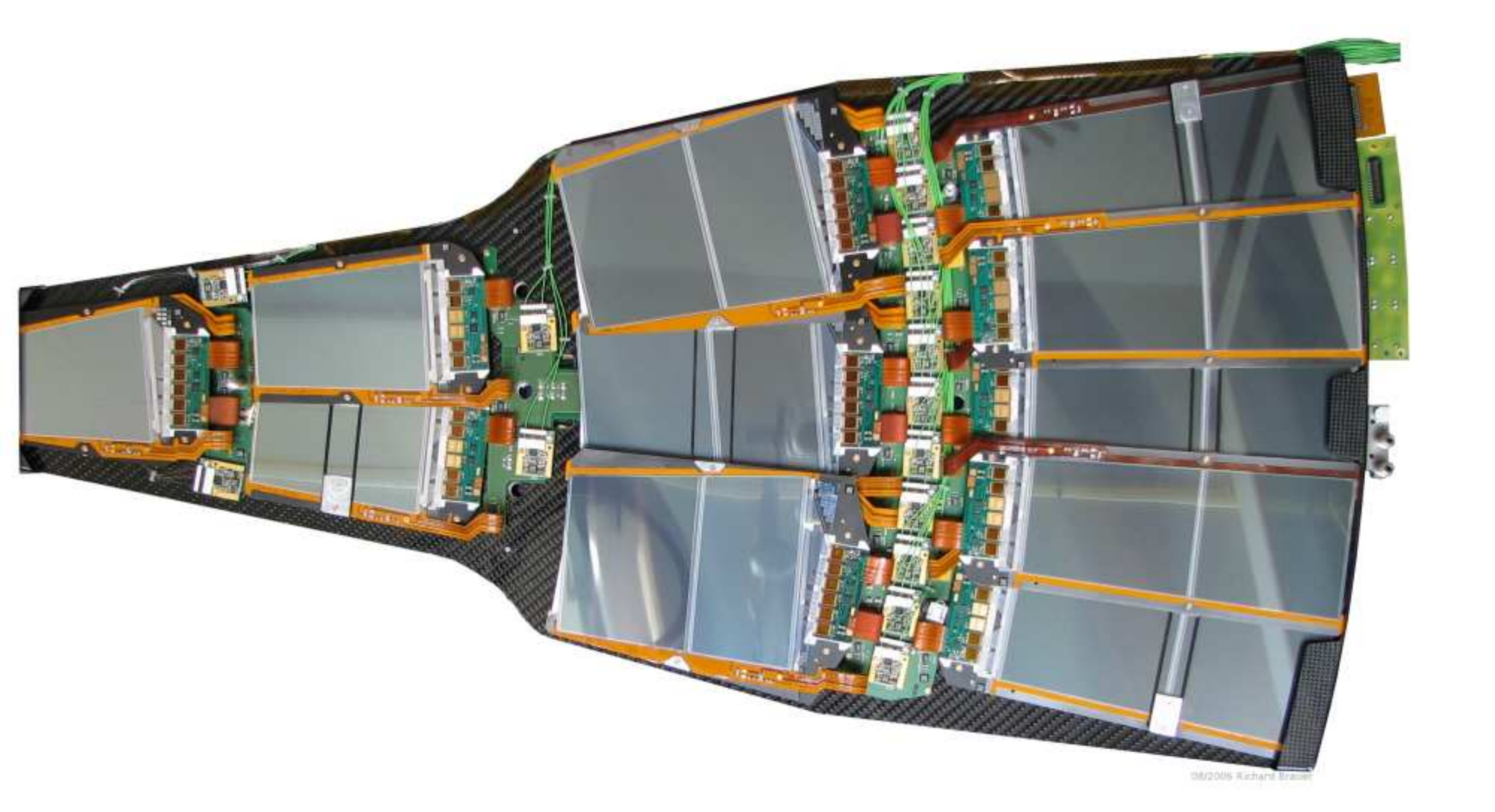}}
    \end{minipage}
    \caption{A TEC front petal (left) and a back petal
    (right) \cite{endcap-integration}. Only detector units on the side
    facing the interaction point can be seen (rings 1,\,3,\,5, and 7). Detector
    units on rings 2,\,4, and 6 are located on the back side of the petal.} 
    \label{fig:application:cmstracker:tec:photopetals}
  \end{center}
\end{figure}

The Tracker Endcap TEC+ covers the range $0.9 \le \eta \le 2.5$ and consists
of 3200 trapezoidal silicon-strip detectors. Both TECs consist of nine disks
(Fig.~\ref{fig:application:cmstracker:tec:endcapdisk}) carrying 16
substructures called petals. Eight petals are mounted on the side facing the
interaction point (front petals) and eight on the far side (back petals). By
grouping the neighbouring front and back petals, the disks are divided into
eight sectors numbered from 1 to 8. On the
petals~(Fig.~\ref{fig:application:cmstracker:tec:photopetals}), detectors are
mounted with the strips in radial direction in up to seven rings. 


\subsection{Sector tests}
\label{sec:application:sectortests}

Petals in TEC+ were integrated and commissioned sector by sector. The petals
were mounted and subsequently attached to services (cooling, power, trigger,
communication and data lines). Connections were tested with custom hard- and
software. Noise data were recorded in order to spot potential flaws, allowing
for repair or replacement. As a final test, signals from cosmic muons
originating from air-showers and traversing the detector were recorded and
used for various studies, including
alignment~\cite{endcap-integration,sprenger}. 

\begin{figure}[ht]
  \begin{center}
    \includegraphics[width=0.48\textwidth]{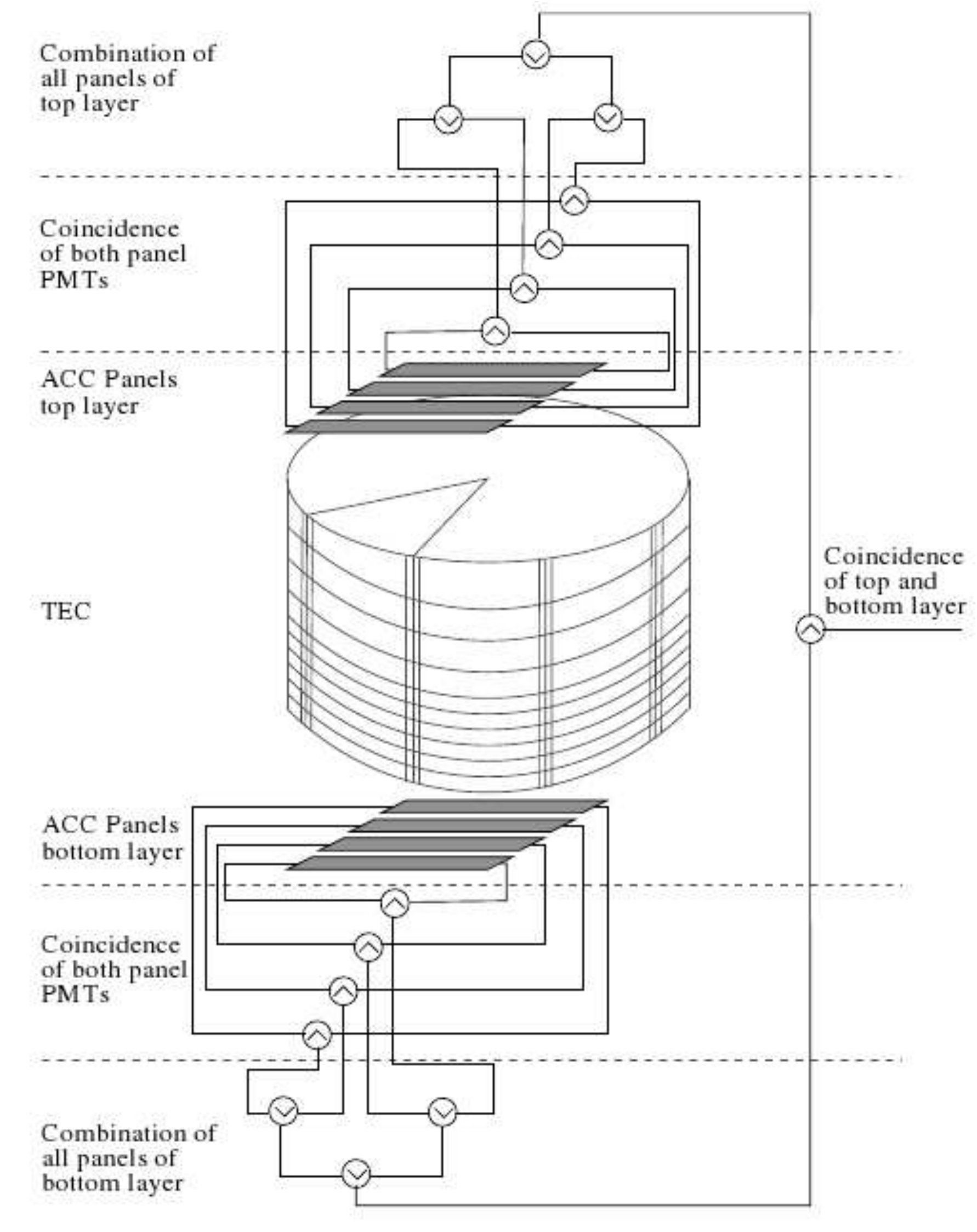}
    \includegraphics[width=0.48\textwidth]{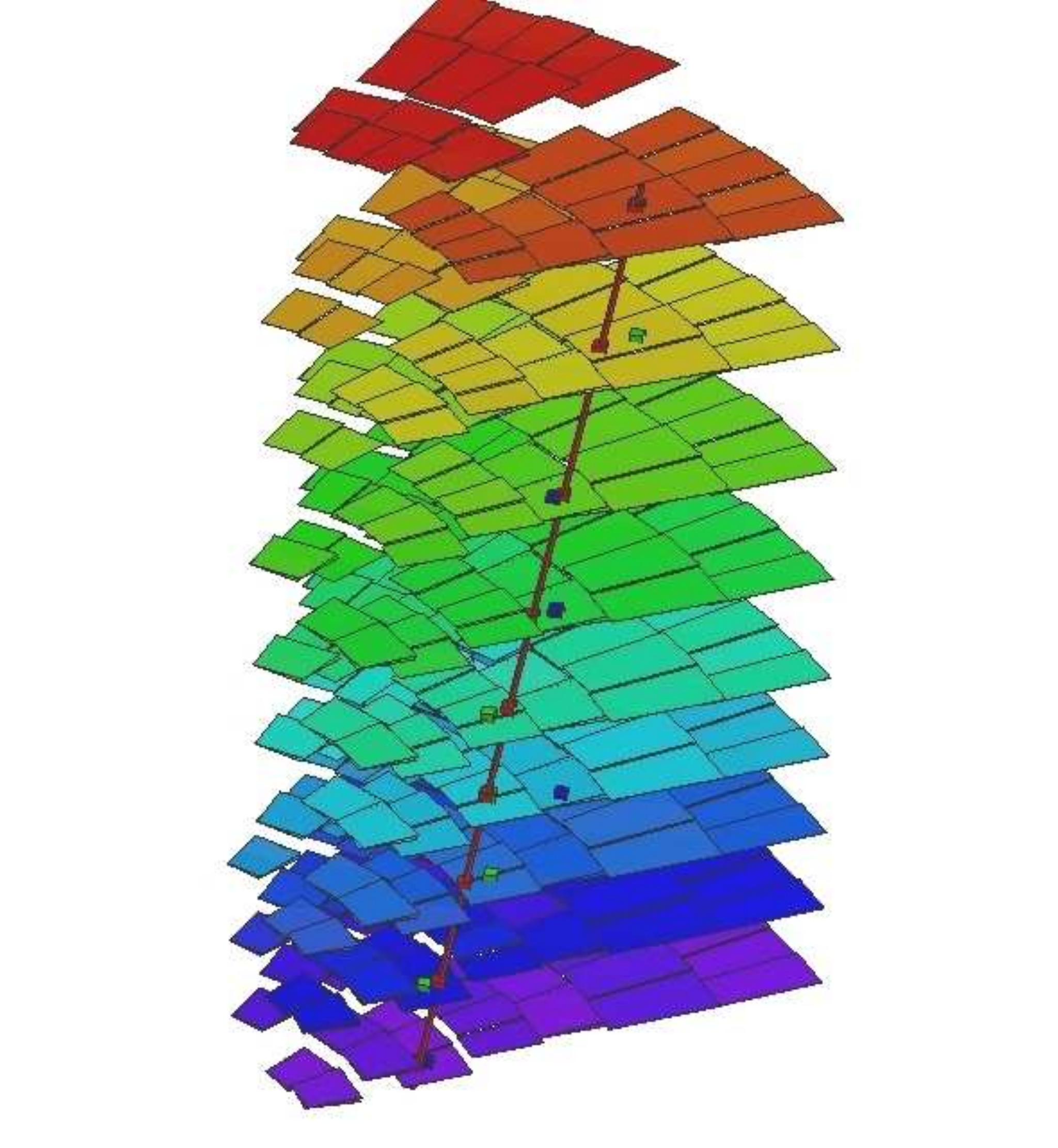}
    \caption{Cosmic trigger setup during TEC+ sector
    tests~\cite{endcap-integration} (left). Reconstructed track of a
    muon traversing the nine disks of the TEC+ (right). Hits used by the track fitting algorithm are displayed in green, all remaining hits are coloured blue.}
    \label{fig:application:sectortests:triggersetup}
  \end{center}
\end{figure}

Figure~\ref{fig:application:sectortests:triggersetup} shows the experimental
setup. During cosmic muon data taking, TEC+ was in a vertical position (disk
planes horizontal). Two square areas of size $80\,\cm\times80\,\cm$, one below
and one above the TEC+ sector under study, were covered with four
AMS~\cite{ams-experiment} scintillator panels~\cite{ams-scintillator},
equipped with two photo-multipliers on opposite sides. A coincidence signal from one
of the upper scintillator panels and one of the lower panels was demanded to
trigger a readout of the TEC+ sector. A $10\,\cm$ thick lead shield was placed
below TEC+, but above the lower scintillators, in order to absorb low energy
($<250\,\MeV$) muons and prevent them from triggering readout. 

Customized CMS software was used to read out the detector, detect signals, and
reconstruct tracks. Table~\ref{tab:application:sectortests:datasets} shows the
number of triggers and reconstructed tracks used for alignment for each sector
in chronological order.


\begin{table}[hb]
  \begin{center}
    \caption{Sector test data sets in chronological order.}
    \label{tab:application:sectortests:datasets}
    \vspace{0.25cm}
    \footnotesize
    \begin{tabular}{|c||c|c|c|}
      \hline
      Run           & Sector  & Number of  & Number of\\
      number        &         & triggers   & tracks\\
      \hline
      \hline
      20944 - 20952 & 5  &  87400 &  6961\\
      21136 - 21163 & 3  & 122058 &  5821\\
      21238 - 21269 & 1  &  88591 &  5807\\
      21428 - 21448 & 7  &  83772 &  8502\\
      21512 - 21530 & 2  &  82139 & 62650*\\
      21592 - 21617 & 6  &  87258 & 68210*\\
      21666 - 21670 & 4  &  58248 & 42260*\\
      21713         & 8  &  73227 & 61159*\\
      \hline
    \end{tabular}\\
    *optimized trigger configuration
  \end{center}
\end{table}

\subsection{Results}
\label{sec:application:results}


\subsubsection{Track selection}

Tracks were reconstructed with bloated hit uncertainties in order to allow an
efficient track reconstruction in the presence of misalignment and selected
with loose criteria. To account for the overestimate, tracks with a
$\chi^2$-probability $P(\tilde\chi^2, n_{dof}) < 0.5$ with $\tilde\chi^2$ from
Eq.~(\ref{eqn:implementation:chi2}) were rejected.

After this selection, 4000 to 6000 tracks were available for each of the
sectors 1,\,3,\,5 and 7. Due to an optimized trigger configuration, which took
place only after integration of the odd-numbered sectors, there were between
35\,000 and 60\,000 tracks available for the even-numbered sectors 2,\,4,\,6
and 8. In total, the sector test data contain about 220\,000 tracks which were
used in alignment studies.

The starting geometry for the linearization of
Eq.~(\ref{eqn:kalmanalgorithm:linearization}) was the design geometry, and the
resulting measurements, predictions, derivatives, and covariance matrices were
stored.

\subsubsection{Disk alignment}
\label{sec:application:results:diskalignmentinxyphi}

Higher level structures like TEC petals and disks were aligned. Here, we describe
the alignment of TEC disks, for which determined alignment parameters
can be compared directly to measurements from survey and the Laser Alignment
System (LAS).

Each of the eight TEC+ sectors has been tested separately. Therefore, no
information can be gathered from the collected data about the relative
position of the sectors with respect to each other. TEC+ disks were aligned by
using all available tracks and assuming that detector elements were mounted on
their nominal position on the petals, and petals on their nominal position on
the disks. For each disk, corrections $\Delta x$, $\Delta y$, and $\Delta\phi$
were calculated. The corrections to the geometry have the same absolute value as
the alignment parameters in Eq.~(\ref{eqn:kalmanalgorithm:updateparameters}),
but with inverted sign.

The Kalman Filter alignment algorithm was configured to use zero as the
initial value of the alignment parameters, corresponding to design
geometry. The covariance matrix was initialised with large startup errors of
$10\,\cm$ and $10\,\rad$ for spatial and angular parameters, respectively, to
avoid biasing the alignment results by the starting values. The reference
system was defined by assigning very small initial uncertainties of
$10^{-5}\,\cm$ and $10^{-5}\,\rad$ to the $x$, $y$, and $\phi$ parameters,
respectively, for the disk closest to the interaction region. A standard
annealing configuration with $b=10\,000$ and $n=100$ was used. Tracks which
resulted in an update of the parameters larger than the parameter uncertainty
were rejected as outliers.

To verify convergence, several starting points for the parameters were tried
and the order in which the data were processed was changed from chronological
to random order. In both cases no significant changes in the resulting alignment
parameters was observed.

Due to the low number of aligned parameters, the
time used for computing the alignment constants, roughly $300\,\second$, was
dominated by input/output operation of the input data. 

Due to the nature of the problem, the objective function only changes
minimally when the linear transformations
\begin{eqnarray*}
  \Delta    x(z) & = & \Delta    x'(z) + a_x    + b_x \cdot z\\
  \Delta    y(z) & = & \Delta    y'(z) + a_y    + b_y \cdot z\\
  \Delta \phi(z) & = & \Delta \phi'(z) + a_\phi + b_\phi \cdot z
\end{eqnarray*}
are applied to the parameters, where $z$ is the coordinate shown in
Figure~\ref{fig:application:cmstracker:cmstracker}.
These transformations correspond to weakly
determined deformations of the TEC+ structure. The computed alignment
parameters were, after the alignment procedure ended, transformed such that

\begin{eqnarray*}
  \sum_i \Delta x_i \cdot z_i & = & 0\\
  \sum_i \Delta x_i           & = & 0
\end{eqnarray*}

and corresponding constraints for the $\Delta y_i$ and $\Delta \phi_i$ were
fulfilled.



\begin{figure}[ht]
  \begin{center}
    \begin{minipage}{7.0cm}
      \centerline{\includegraphics[width=7.0cm]{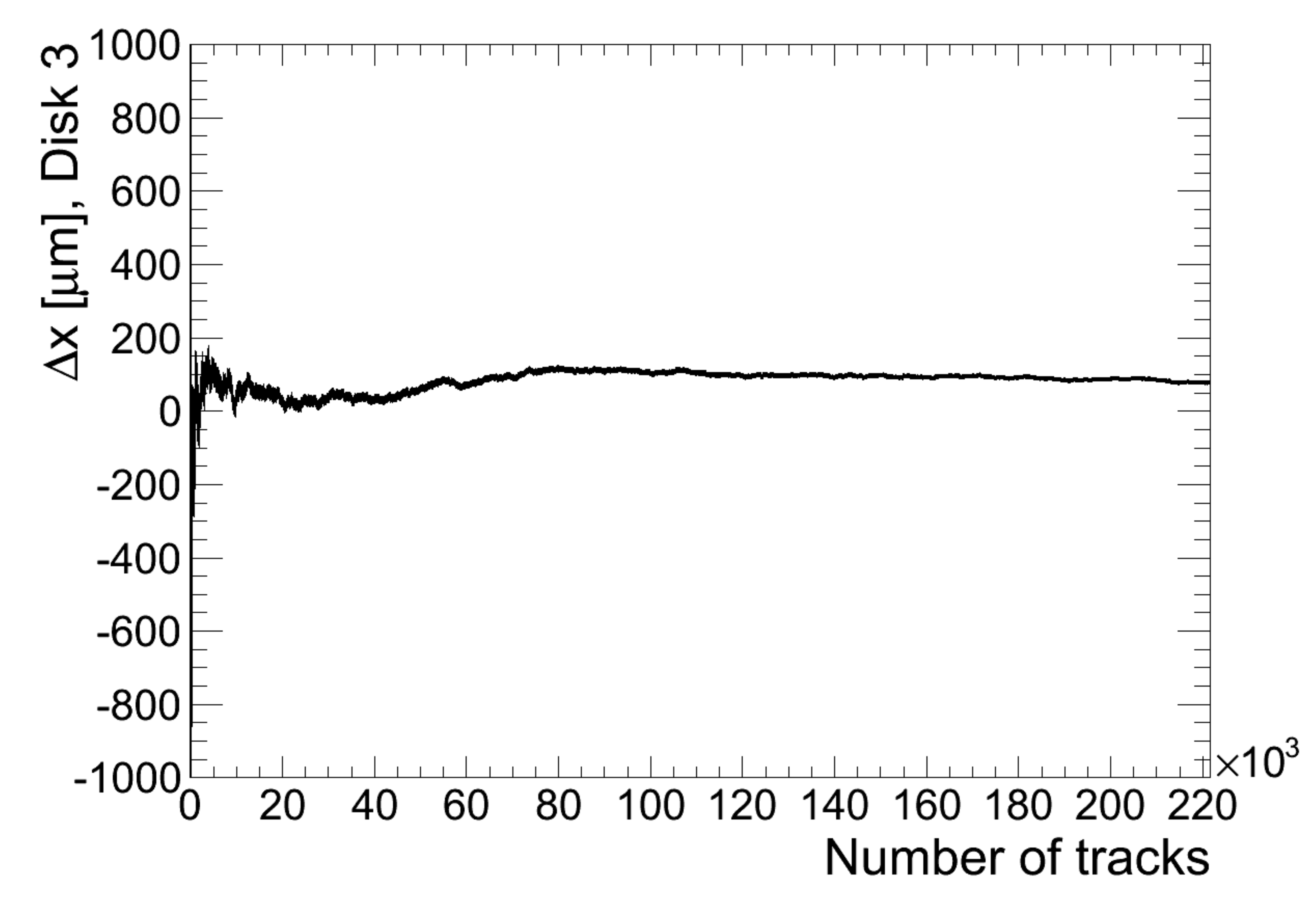}}
    \end{minipage}
    \hspace{0.6cm}
    \begin{minipage}{7.0cm}
      \centerline{\includegraphics[width=7.0cm]{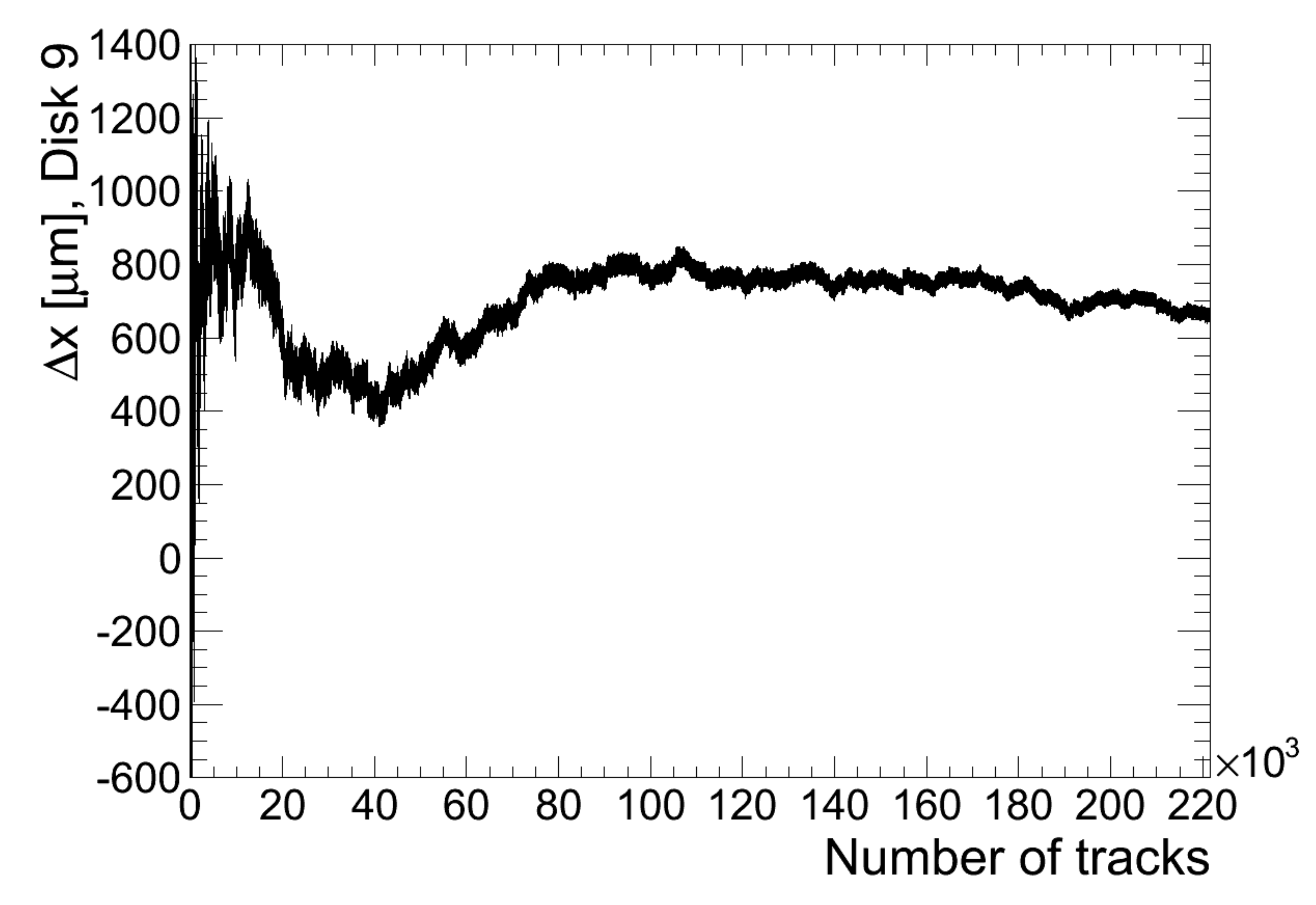}}
    \end{minipage}
    \caption{Corrections $\Delta x$ for disks 3 (left) and 9 (right) as a
    function of the number of processed tracks. The width of the band
    represents the uncertainty on the correction.}
    \label{fig:application:results:acdisksxyg-evolution}
  \end{center}
\end{figure}

Figure~\ref{fig:application:results:acdisksxyg-evolution} shows the
corrections $\Delta x$ for disks 3 and 9 as a function of the number of
processed tracks. It can be seen that the uncertainty on the
correction decreases with the number of processed tracks. The uncertainty is larger
for corrections belonging to alignables which are farther away from the disk that
is used as the reference system (disk~1).

\begin{figure}[ht]
  \begin{center}
    \begin{minipage}{7.0cm}
      \centerline{\includegraphics[width=7.0cm]{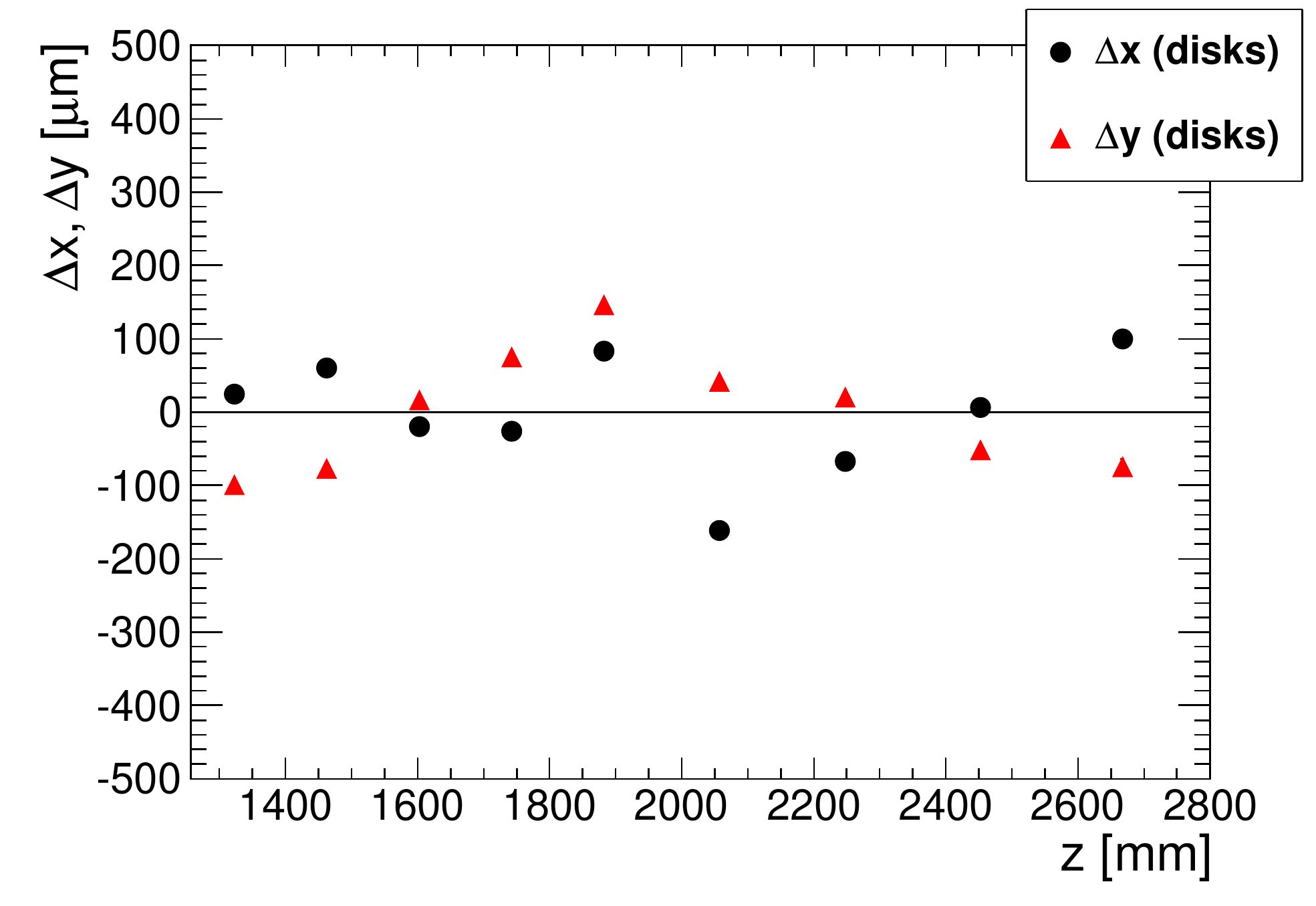}}
    \end{minipage}
    \hspace{0.6cm}
    \begin{minipage}{7.0cm}
      \centerline{\includegraphics[width=7.0cm]{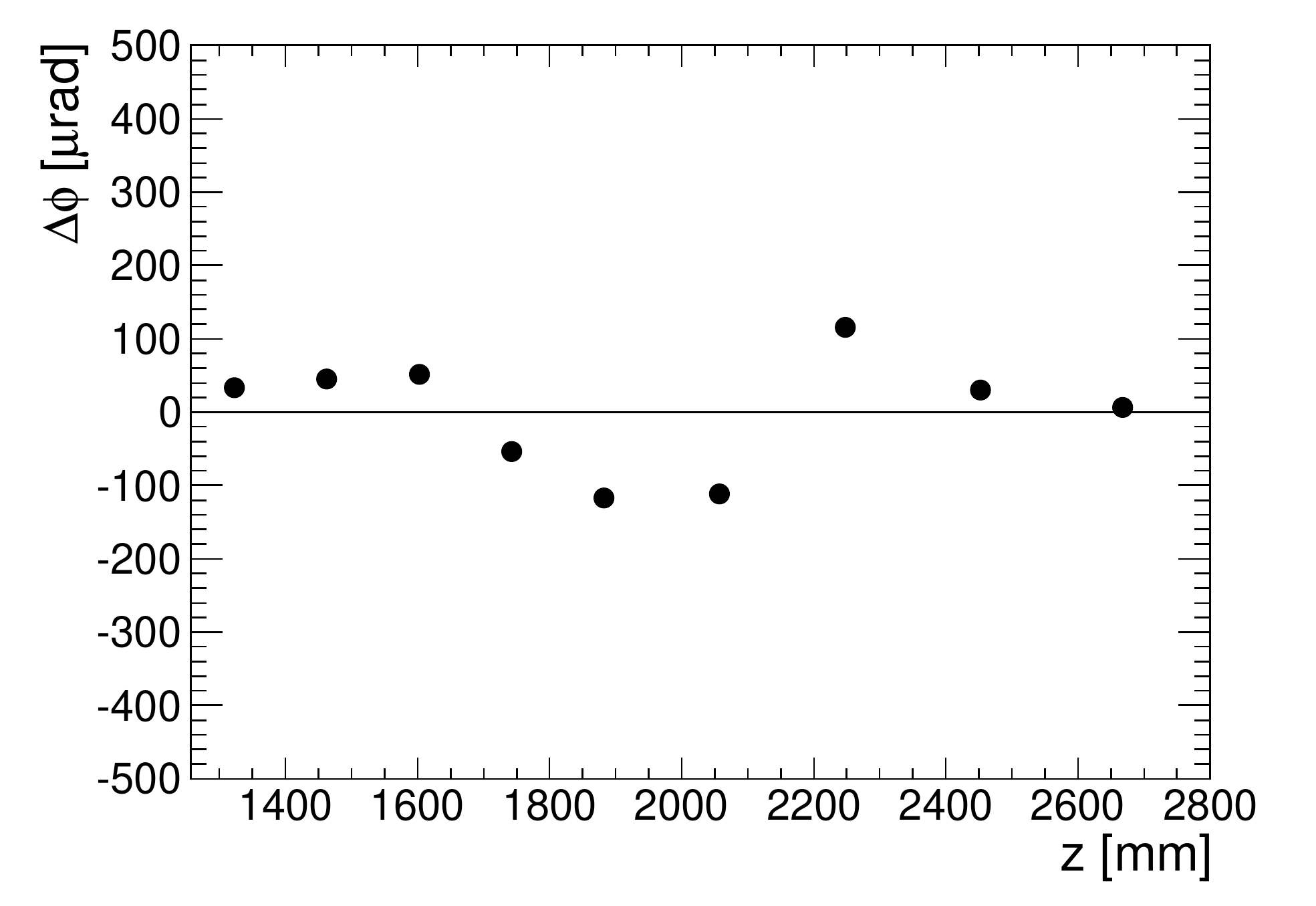}}
    \end{minipage}
    \caption{Alignment corrections for disk alignment in $x, y, \phi$. Error
    bars are smaller than the symbol size.}
    \label{fig:application:results:acdisksxyg-limits}
  \end{center}
\end{figure}

Table~\ref{tab:sectortestdata:acdisksxyg} lists the obtained alignment
corrections together with their errors. The result is also displayed
graphically in Figure~\ref{fig:application:results:acdisksxyg-limits}. The
size of the corrections is of order $100\,\mum$ in $x$ and $y$, and
$100\,\micro\rad$ in $\phi$, which corresponds to an arc length of about
$100\,\mum$ at the outer disk circumference. The accuracy of the positional
and rotational corrections, $2$--$10\,\mum$ and $2$--$10\,\micro\rad$,
respectively, is very high compared to the typical size of the correction.

\begin{table}[hb]
  \begin{center}
    \caption{Alignment corrections obtained from disk alignment in $x$, $y$, $\phi$}
    \label{tab:sectortestdata:acdisksxyg}
    \vspace{0.25cm}
    \footnotesize
    \begin{tabular}{|c||r|r|r|}
      \hline
      Disk number	 & \multicolumn{1}{c|}{$\Delta x\,[\mum]$}	 & \multicolumn{1}{c|}{$\Delta y\,[\mum]$}	 & \multicolumn{1}{c|}{$\Delta\phi\,[\micro\rad]$}	 \\ \hline \hline
      1	 & 24 $\pm$ 8	 & -99 $\pm$ 8	 & 33.6 $\pm$ 7.3	 \\
      2	 & 60 $\pm$ 6	 & -77 $\pm$ 6	 & 45.3 $\pm$ 5.6	 \\
      3	 & -20 $\pm$ 4	 & 17 $\pm$ 4	 & 51.2 $\pm$ 4.1	 \\
      4	 & -26 $\pm$ 3	 & 76 $\pm$ 3	 & -53.6 $\pm$ 2.7	 \\
      5	 & 83 $\pm$ 2	 & 146 $\pm$ 2	 & -117.0 $\pm$ 1.7	 \\
      6	 & -162 $\pm$ 2	 & 42 $\pm$ 2	 & -111.4 $\pm$ 1.9	 \\
      7	 & -67 $\pm$ 4	 & 21 $\pm$ 4	 & 115.8 $\pm$ 3.7	 \\
      8	 & 7 $\pm$ 6	 & -51 $\pm$ 7	 & 29.9 $\pm$ 6.2	 \\
      9	 & 100 $\pm$ 10	 & -74 $\pm$ 10	 & 6.2 $\pm$ 9.2	 \\
      \hline
    \end{tabular}
  \end{center}
\end{table}

\subsubsection{Comparison to Laser Alignment System results}
\label{sec:application:results:lascomparison}

\begin{figure}[ht]
  \begin{center}
    \begin{minipage}{7.0cm}
      \centerline{\includegraphics[width=7.0cm]{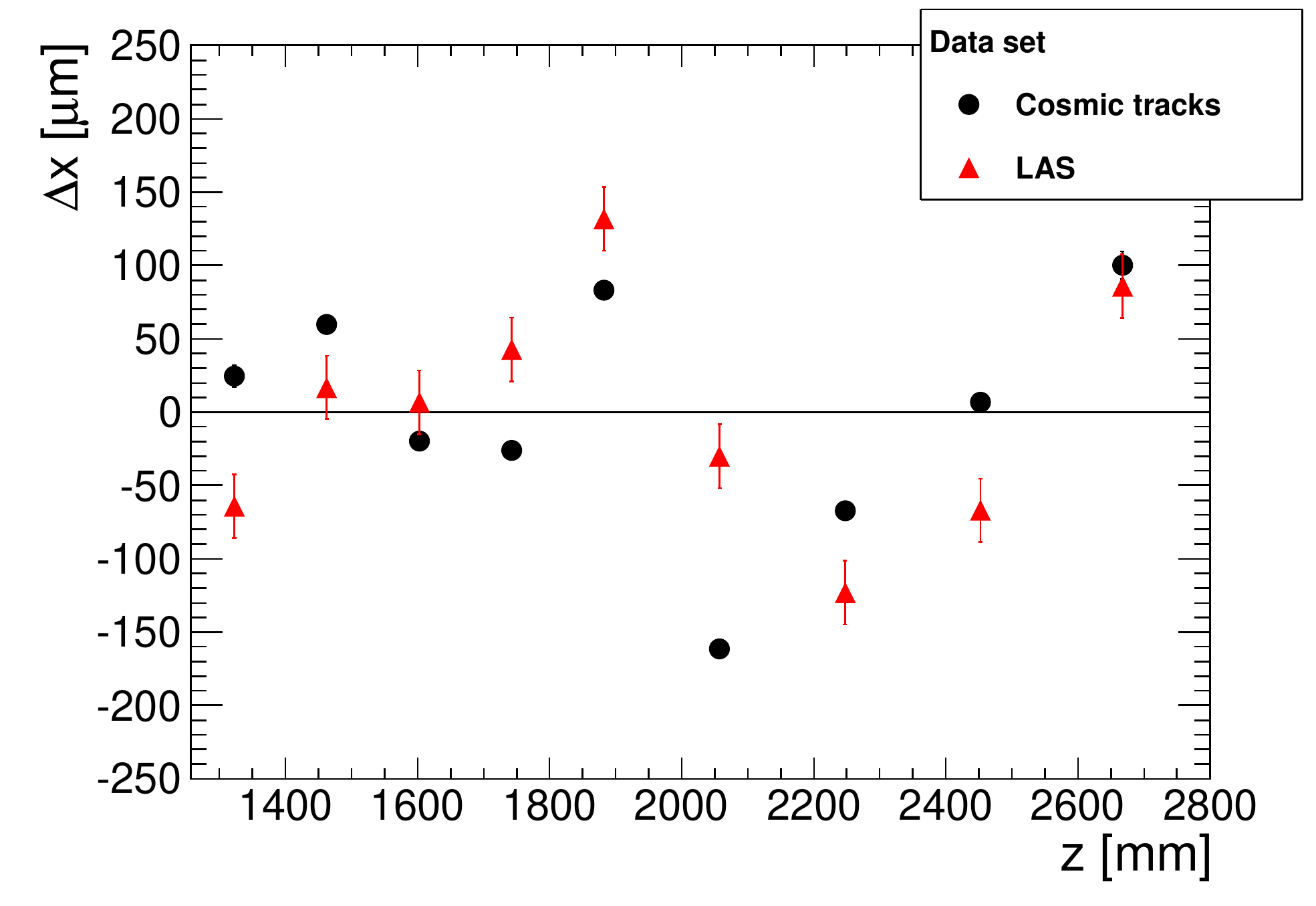}}
    \end{minipage}
    \hspace{0.6cm}
    \begin{minipage}{7.0cm}
      \centerline{\includegraphics[width=7.0cm]{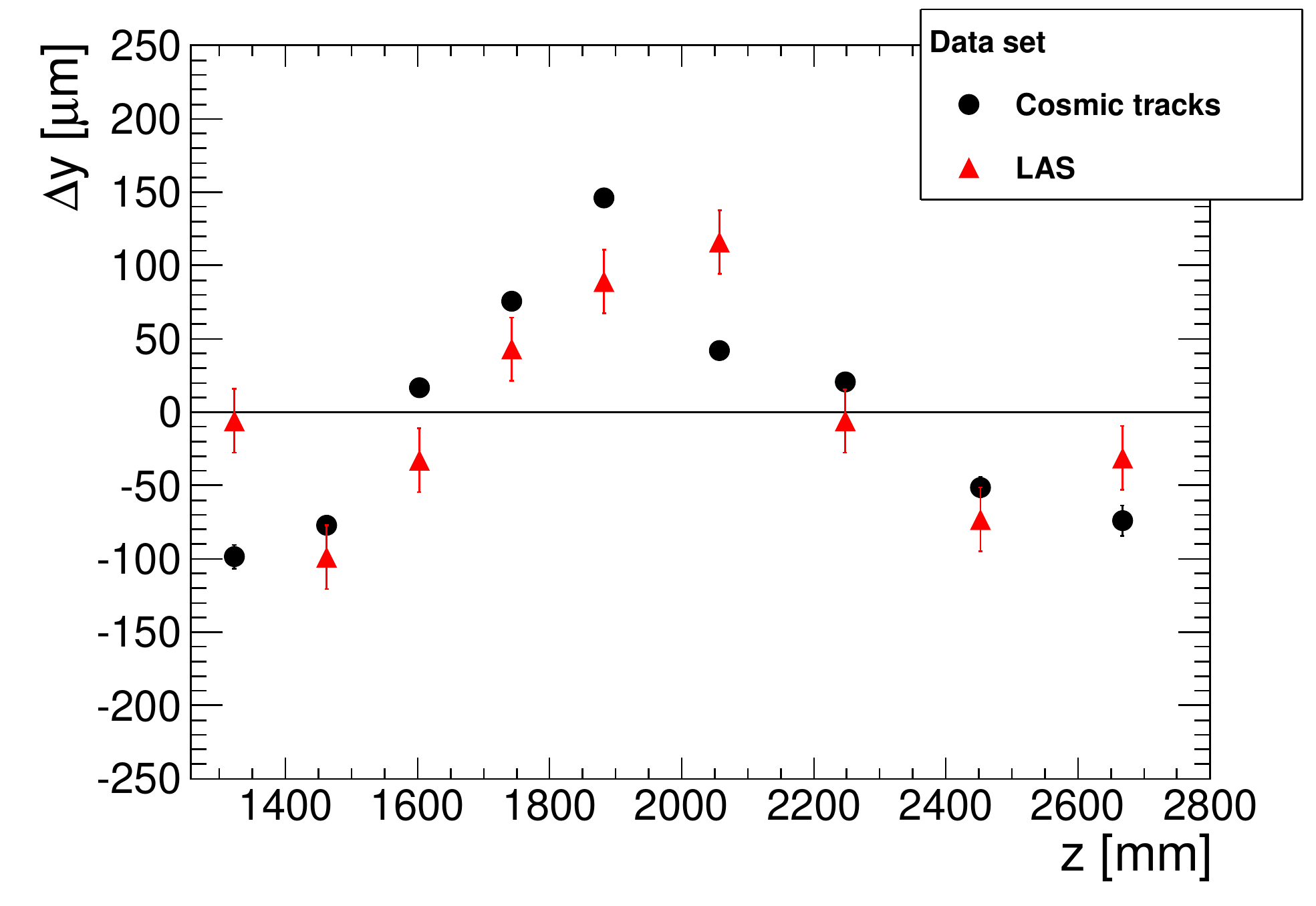}}
    \end{minipage}\\
    \vspace{0.6cm}
    \begin{minipage}{7.0cm}
      \centerline{\includegraphics[width=7.0cm]{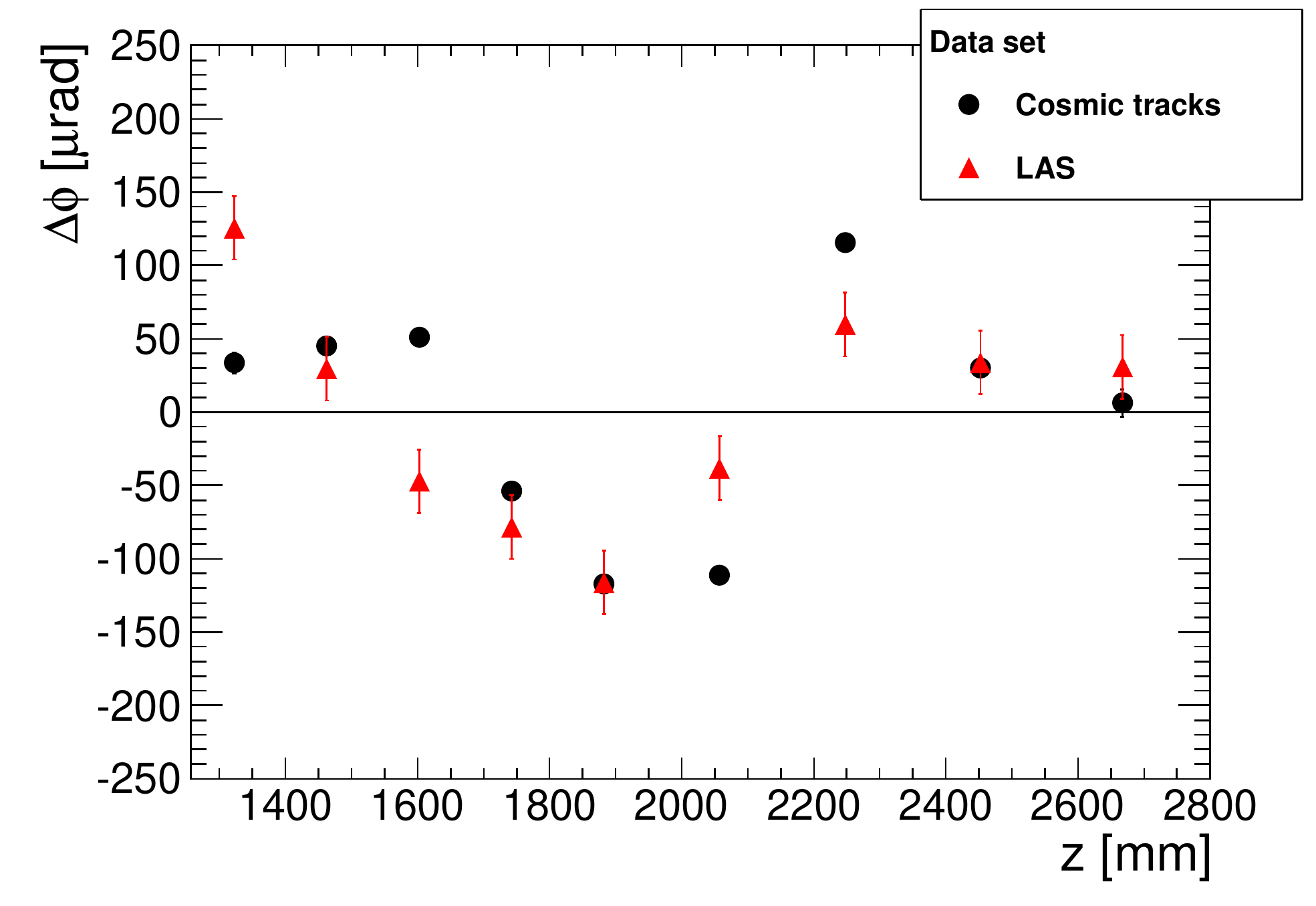}}
    \end{minipage}
    \caption{Comparison of Laser Alignment System \cite{vci}\
    and track-based alignment results.} 
    \label{fig:application:results:lascomparison:aachendiscs-limits}
  \end{center}
\end{figure}

The TEC also comprises a hardware alignment system~\cite{cms}, using sixteen laser rays
of wavelength $\lambda = 1075\,\nm$ propagating parallel to the beam line. All
rays pass through the back petals, eight rays pass at a radial distance $r_4 =
564\,\mm$ and eight further rays at $r_6 = 840\,\mm$, the subscript indicating
the ring number.  The rays pass through an opening in the silicon detector's
backside metallization. The laser rays get partially absorbed and produce
signal on the detector elements. Using a straight line hypothesis, residuals
can be deduced. These residuals were used to determine corrections to position and
rotation of the nine TEC disks, $\Delta x$, $\Delta y$, and $\Delta \phi$~\cite{vci}. The corrections are
listed in Table~\ref{tab:resultinterpretation:aclasdiskxyg} and compared to
the results deduced from the Kalman Filter in
Figure~\ref{fig:application:results:lascomparison:aachendiscs-limits}.

\begin{table}[hb]
  \begin{center}
    \caption{Displacements $\Delta x$, $\Delta y$, $\Delta \phi$ of TEC+ disks
      determined with LAS residuals. The precision of the position corrections
      $\Delta x$ and $\Delta y$ is $23\,\mum$, and $\Delta \phi$ is
      determined with an accuracy of $23\,\micro\rad$.}
    \label{tab:resultinterpretation:aclasdiskxyg}
    \vspace{0.25cm}
    \footnotesize
    \begin{tabular}{|c||r|r|r|}
      \hline
      Disk number	 & \multicolumn{1}{c|}{$\Delta x\,[\mum]$}	 & \multicolumn{1}{c|}{$\Delta y\,[\mum]$}	 & \multicolumn{1}{c|}{$\Delta\phi\,[\micro\rad]$}	 \\ \hline \hline
      1	 & -64	 & -6	 & 126	 \\
      2	 & 17	 & -99	 & 30	 \\
      3	 & 7	 & -33	 & -47	 \\
      4	 & 43	 & 43	 & -78	 \\
      5	 & 132	 & 89	 & -116	 \\
      6	 & -30	 & 116	 & -38	 \\
      7	 & -123	 & -6	 & 60	 \\
      8	 & -67	 & -73	 & 34	 \\
      9	 & 86	 & -31	 & 31	 \\
      \hline
    \end{tabular}
  \end{center}
\end{table}

The RMS of the differences between Laser Alignment System and Kalman Filter
corrections is $70\,\mum$ in $\Delta x$, $52\,\mum$ in $\Delta y$, and
$56\,\micro\rad$ in $\Delta \phi$. Although overall good agreement is
observed, some measurements deviate significantly. However, perfect agreement
is not expected since the Laser Alignment System relies exclusively on 144
measurements of detector elements in rings 4 and 6 of back petals only,
whereas the track based alignment comprises data from more than 3000 modules
on both front and back petals. In a further analysis it was found that the
differences are compatible with the mounting precision of the
petals. Especially the large deviation in $\Delta x$ and $\Delta y$ for the
disk at $z \approx 2050\,\mm$ could be traced back to an imperfect front petal
mounting in sector eight. Here, one of the precision pins with which the
petals are attached to the disks was visually confirmed to be bent into the
direction indicated by the alignment corrections.

Interpreting the result, the largest contribution to the misalignment is due
to the positioning of the disks in the TEC+, and the remaining difference is
to a large extent due to the positioning of the petals on the disks. This is
in agreement with the na\"{\i}ve expectation that smaller structures can be
assembled and mounted with higher precision than larger structures.

\subsubsection{Comparison to survey measurements}
\label{sec:application:results:surveycomparison}

A further validation of the alignment results was performed with survey data
recorded by a survey team before integrating the petals and after their
integration. Using photogrammetry, the displacements of four points at the
outer circumference of each disk were measured with a precision of about
$60\,\mum$. Each measurement was done once with the TEC+ being in horizontal
and once with it being in vertical orientation. As before, corrections $\Delta
x$ and $\Delta y$ to the disk position as well as $\Delta \phi$ to the disk
rotation were estimated from these
measurements. 

\begin{figure}[ht]
  \begin{center}
    \begin{minipage}{7.0cm}
      \centerline{\includegraphics[width=7.0cm]{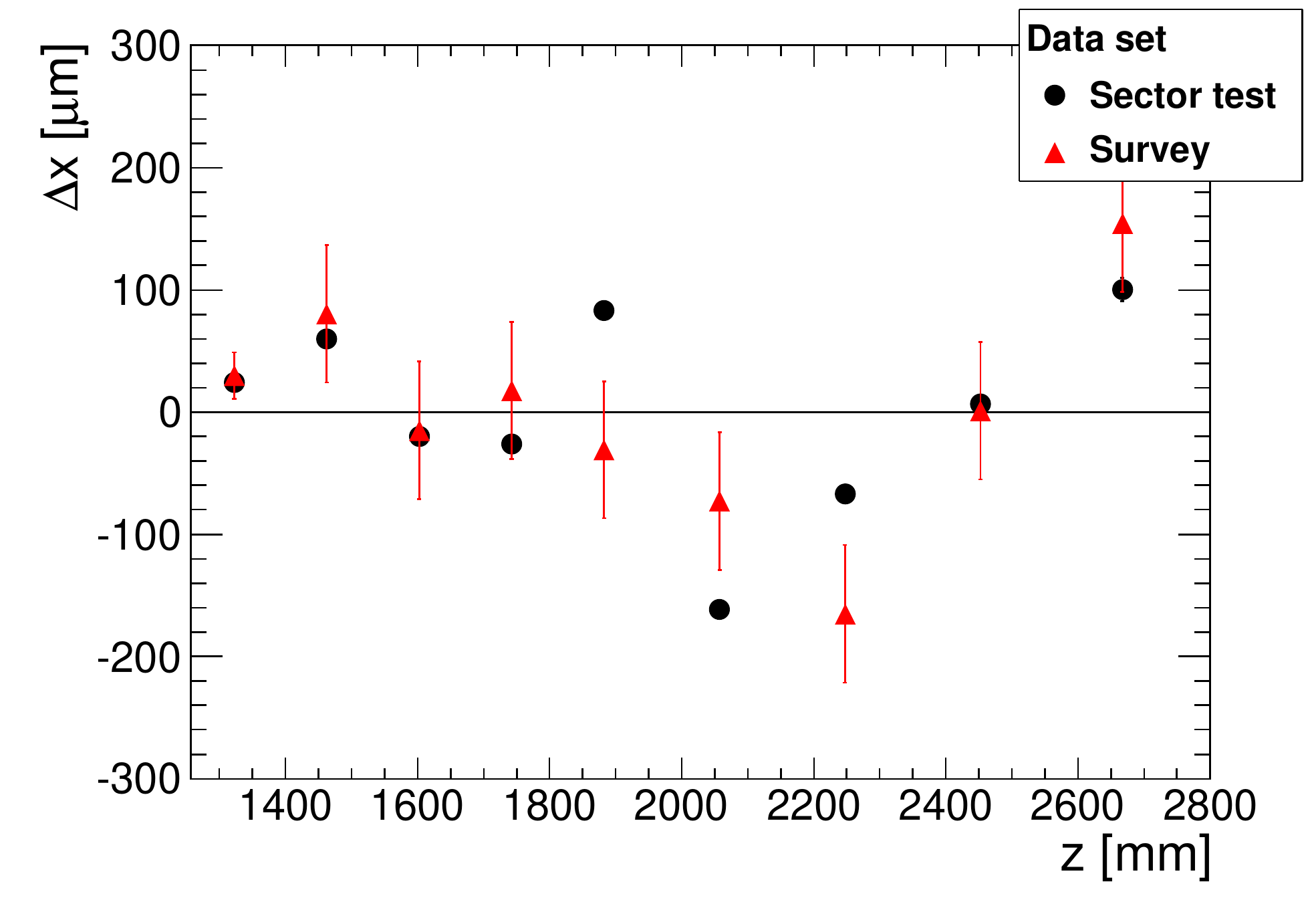}}
    \end{minipage}
    \hspace{0.6cm}
    \begin{minipage}{7.0cm}
      \centerline{\includegraphics[width=7.0cm]{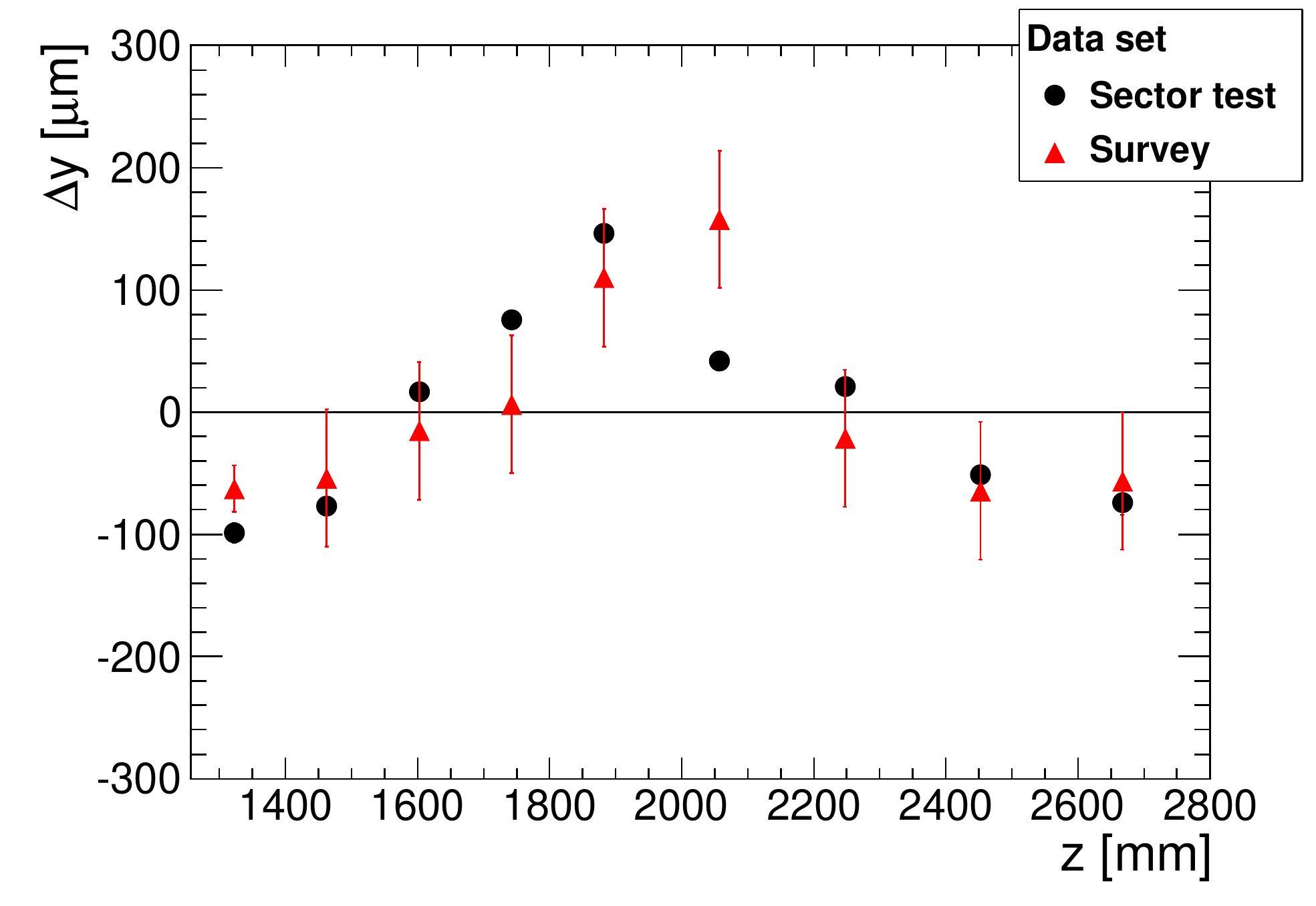}}
    \end{minipage}\\
    \vspace{0.6cm}
    \begin{minipage}{7.0cm}
      \centerline{\includegraphics[width=7.0cm]{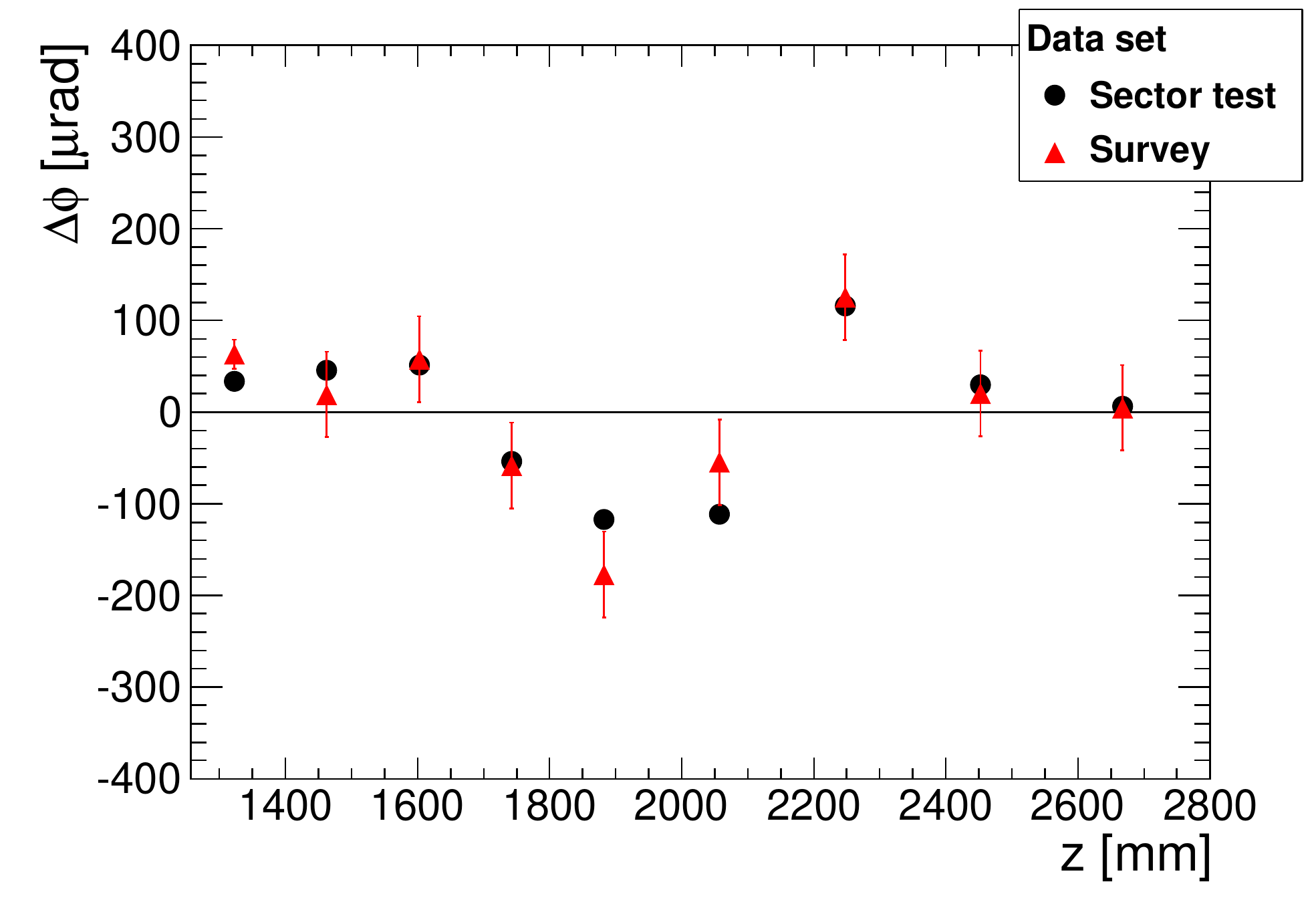}}
    \end{minipage}
    \caption{Comparison of survey 
    and track-based alignment results.}
    \label{fig:application:results:surveycomparison:limits}
  \end{center}
\end{figure}

The measurements were transformed into the same reference system as used for
the track-based alignment and are listed in
Table~\ref{tab:resultinterpretation:acsurveyxyg}. A graphical comparison of
the track-based alignment results and survey data is given in
Figure~\ref{fig:application:results:surveycomparison:limits}. The track-based
alignment results agree with the survey measurements within (RMS) $63\,\mum$
in $\Delta x$, $52\,\mum$ in $\Delta y$, and $34\,\micro\rad$ in $\Delta \phi$
and show, within measurement precision, good agreement.

\begin{table}[htb]
  \begin{center}
    \caption{Displacements $\Delta x$, $\Delta y$, $\Delta \phi$ of TEC+ disks determined with survey measurements. The precision of the position corrections $\Delta x$ and $\Delta y$ is $57\,\mum$, and $\Delta \phi$ is determined with an accuracy of $47\,\micro\rad$.}
    \label{tab:resultinterpretation:acsurveyxyg}
    \vspace{0.25cm}
    \footnotesize
    \begin{tabular}{|c||r|r|r|}
      \hline
      Disk number	 & \multicolumn{1}{c|}{$\Delta x\,[\mum]$}	 & \multicolumn{1}{c|}{$\Delta y\,[\mum]$}	 & \multicolumn{1}{c|}{$\Delta\phi\,[\micro\rad]$}	 \\ \hline \hline
      1  & 30     & -63    & 63\\
      2  & 81     & -54    & 19\\
      3  & -15    & -15    & 57\\
      4  & 18     & 6     & -5\\
      5  & -31    & 110     & -177\\
      6  & -73    & 158     & -55\\
      7  & -165    & -21    & 125\\
      8  & 1     & -64    & 20\\
      9  & 154     & -56    & 5\\
      \hline
    \end{tabular}
  \end{center}
\end{table}

\section{Summary}
\label{sec:summary}

A Kalman Filter alignment algorithm has been implemented in an
experiment-independent way and applied to data. The obtained alignment
corrections have been validated with both a hardware alignment system and
survey data. The difference (RMS) between the corrections estimated with the
track-based alignment and the hardware alignment system is $70\,\mum$ and
$52\,\mum$ in $\Delta x$ and $\Delta y$, respectively, and $56\,\micro\rad$ in
$\Delta \phi$. The track-based alignment results agree with the survey
measurements within (RMS) $63\,\mum$ in $\Delta x$, $52\,\mum$ in $\Delta y$,
and $34\,\micro\rad$ in $\Delta \phi$. The difference betweeen laser alignment
and survey data consistently is of a similar magnitude. Statistical
uncertainties on the obtained alignment corrections are negligible when
compared to the statistical uncertainties of laser alignment and survey data.


\section*{Acknowledgements}

We would like to thank Rudi Fr\"uhwirth, Wolfgang Adam and Edmund Widl for
giving us insight into the Kalman Filter alignment algorithm. 


\end{document}